\documentclass[11pt,oneside,english]{article}
\usepackage[a4paper, total={6.2in, 9.5in}]{geometry}
\usepackage[utf8]{inputenc}
\pdfoutput=1

\usepackage{rotating}
\usepackage{graphicx}
\usepackage{amssymb}
\usepackage[table]{xcolor}
\usepackage{xcolor}
\usepackage[T1]{fontenc}
\usepackage{lmodern}

\usepackage{multirow,booktabs,array,longtable,caption}

\usepackage{pdflscape}

\definecolor{verylightgray}{gray}{0.9}

\newcommand{\changess}[1]{#1}
\newcommand{\changes}[1]{#1}

\usepackage{enumitem}
\newlist{compactimize}{itemize}{1}
\setlist[compactimize]{label=\textbullet, nosep, left=0pt,
                    before={\begin{minipage}[t]{\hsize}},
                    after ={\end{minipage}} }

\usepackage{caption}

\usepackage{courier}  

\usepackage[bottom, multiple]{footmisc}  
\usepackage[normalem]{ulem}    % for \sout (line-through)
\usepackage{anyfontsize}

\PassOptionsToPackage{hyphens}{url}\usepackage[hidelinks, hyperfootnotes=false]{hyperref}
\usepackage{fancyhdr}                           % Needed for footers on each page
\fancypagestyle{firstpage} % first page footer only
{
   \footskip=8mm
   \fancyhf{}
   \fancyfoot[C]{\vspace{-7mm}\thepage}

   \fancyfoot[L]{\vspace{-1mm}\noindent\rule{6.2in}{0.4pt}
     \scriptsize{ 
       *This paper appeared in the Journal of Financial Market Infrastructures,
       Vol 12, No 1 (December 2024), pages 43-71. \newline
       \textcopyright \hspace{0.5mm} \changess{Barclays 2023-2024} \\
       This work is licensed under a \href{https://creativecommons.org/licenses/by/4.0/}{Creative Commons Attribution 4.0 International License (CC
       BY).}  \\ Provided you adhere to the CC BY license, including as to attribution, you are
       free to copy and redistribute this work in any medium or format and remix, transform,
       and build upon the work for any purpose, even commercially.                  \\
       BARCLAYS is a registered trademark, all rights are reserved. \\
     }
 }
}
\title{\vspace{-1cm}Functional Consistency across \\ 
Retail Central Bank Digital Currency and \\
Commercial Bank Money*}
 \vspace{1cm}

\author{%
  \begin{tabular}{c} {\fontsize{10.75}{1cm}\selectfont Lee Braine, Shreepad Shukla and Piyush Agrawal} \\ {\fontsize{10.75}{1cm}\selectfont Chief Technology Office} \\
    {\fontsize{10.75}{1cm}\selectfont Barclays} \\ \hskip 1em \end{tabular} }

\date{\vspace{-0.5cm}August 16, 2023}

\begin{document}
\maketitle
\vspace{-1.40cm}

\begin{center}{\fontsize{8}{1cm}\selectfont (Revised Novemeber 1, 2024)}\end{center}
\thispagestyle{firstpage} % Needed to get footer on first page
%\vspace{-0.75cm}

%%%%%%%%%%%%%%%%%%%%%%%%%%%%%%%%%%%%%%%%%%%%%%%%%%%%%%%%%%%%%%%%%%%
%% Section: Abstract
%%%%%%%%%%%%%%%%%%%%%%%%%%%%%%%%%%%%%%%%%%%%%%%%%%%%%%%%%%%%%%%%%%%

\begin{abstract}
\noindent
Central banks are actively exploring central bank
digital currencies (CBDCs) by conducting research, proofs
of concept and pilots.
However, adoption of a retail CBDC can risk fragmenting both
payments markets and retail deposits if the retail CBDC and
commercial bank
money do not have common operational characteristics.
\changess{In this paper we}
focus on a potential UK retail CBDC \changess{\textendash} the ``digital pound'' \changess{\textendash} and
the Bank of England's ``platform model''.
We first explore how the concept of functional consistency
could mitigate the risk of fragmentation.
We next identify the common operational characteristics that are
required to achieve functional consistency across all forms
of regulated retail digital money.
We identify four design options based on the provision of these
common operational characteristics by the central bank, 
\changess{payment interface providers, technical service providers or
a financial market infrastructure.}
We next identify architecturally significant use cases and select
key capabilities that support these use cases and the common
operational characteristics.
We evaluate the suitability of the
design options to provide these key capabilities and draw insights.
We conclude that no single design option could provide
functional consistency across digital pounds and commercial bank 
money and, instead,
a complete solution would need to combine the suitable design
\changess{option(s) for each key capability.}
\end{abstract}

%\vspace{0.15cm}

%%%%%%%%%%%%%%%%%%%%%%%%%%%%%%%%%%%%%%%%%%%%%%%%%%%%%%%%%%%%%%%%%%%
%% Section 1: Introduction
%%%%%%%%%%%%%%%%%%%%%%%%%%%%%%%%%%%%%%%%%%%%%%%%%%%%%%%%%%%%%%%%%%%

\section{Introduction}
\label{sec:introduction}

\noindent
A central bank digital currency (CBDC) is a digital payment instrument,
denominated in a national unit of account, that is a direct liability
of a central bank \cite{bis-aer2021-cbdcs}.
Central banks are actively exploring retail CBDCs 
\changess{\cite{boe-cbdc-mainpage, ecb-digeuro-stocktake, rbi-cbdc-note, bis-cbdc-survey-2023, pboc-ecny-paper}}
with various motivations \changess{such as
improving financial stability,
implementing monetary policy, 
encouraging financial inclusion, 
improving (domestic and cross-border) payment efficiency, and
enhancing payments safety and robustness \cite{bis-cbdc-survey-2023}}.

\changess{The Bank of England identified its primary motivations
for a potential UK retail CBDC, the ``digital pound'', as \cite{boe-cbdc-cons-paper}:
\begin{itemize}
\item sustained access to
central bank money as an anchor for confidence and safety in
the monetary system, and 
\item promotion of innovation and choice in domestic
payments.
\end{itemize}
}
The Bank of England and HM Treasury \changess{established} the CBDC
Taskforce to coordinate the exploration of a potential digital pound, as well as
two external engagement groups, the CBDC Engagement Forum and the CBDC
Technology Forum, to gather input on \linebreak non-technology and
technology aspects respectively of a potential digital pound \cite{boe-cbdc-mainpage}.
\changess{The Bank of England and HM Treasury consultation paper (CP)
\cite{boe-cbdc-cons-paper}
sets out their assessment of the case for a
digital pound, while
the Bank of England technology working paper (TWP)
\cite{boe-cbdc-twp}
outlines its emerging thinking on CBDC
technology.}
\changess{The Bank of England and HM Treasury have published
responses \cite{boe-cbdc-cp-resp} to the CP and responses
\cite{boe-cbdc-twp-resp} to the TWP based on feedback
received.}

The design of a retail CBDC and its underlying system could potentially
lead to significant risks, ranging from cyber security risks \cite{bis-aer2021-cbdcs}
to financial stability risks \cite{bis-gocb-cbdc-finstab}.
\changess{In a previous paper \cite{barc-cbdc-iia-doi}}, where we 
focused on the digital pound,
\changess{we identified the additional risk of fragmentation in payments
markets and retail deposits if digital pounds and commercial bank
money do not have common operational characteristics, and
we presented an illustrative industry architecture intended to
mitigate this risk by placing the digital pound and commercial bank
money on a similar footing.}
We subsequently developed prototypes of this industry
\changess{architecture as part of the} Barclays CBDC Hackathon 2022\footnote
{Barclays CBDC Hackathon 2022 \cite{barclays-cbdchack-site} 
consisted of a series of coding challenges 
that explored the future of money, including both the digital pound
and commercial bank money.}
and Project Rosalind\footnote
{Project Rosalind \cite{bis-rosalind-site} is a joint experiment
on a retail CBDC led by the Bank for International Settlements’
Innovation Hub London Centre and the Bank of England,
in collaboration with the private sector.}.

In this paper we continue to focus on the digital pound and the
risk of fragmentation in payments markets and retail deposits.
We first explore the important concept of ``functional consistency'' 
\changess{(see Section \ref{sec:func-cons})}
across all forms of regulated retail digital money, including 
existing forms of money (such as commercial bank deposits and e-money)
and new forms of money (such as retail CBDCs, tokenised deposits and
regulated stablecoins).
Functional consistency requires several common operational
characteristics and extends
the concept of fungibility so that units of one form of
money can be substituted for units of other forms of money 
(and not just for other units of the same form of money)\changess{; we} believe functional consistency is a desirable principle
that could mitigate the risk of fragmentation in retail deposits
and payments.
We then present design options that are based on the Bank of England's
``platform model'' \cite{boe-cbdc-cons-paper} and \changess{that} aim to support
functional consistency and interoperability across the digital
pound and commercial bank money \changess{(see Section \ref{sec:uk-cbdc-des-opts})}.
We next describe a set of architecturally significant use cases and
the key capabilities they require \changess{(see Section \ref{sec:uk-cbdc-uc-cap})}.
These key capabilities support some of the common operational
characteristics that are needed to achieve functional consistency.
We then present our preliminary evaluation of
the suitability of the presented design options
to support the key capabilities\changess{,} and we draw some initial insights 
\changess{(see Section \ref{sec:uk-cbdc-des-opts-anal}).
Finally, we state our conclusions and }identify potential next steps
\changess{(see Sections \ref{sec:conclusions}
and \ref{sec:summary-and-further-work})}.

\changess{Our analysis focuses on technical and functional
aspects, so further work could include a comprehensive analysis
of legal, operational and commercial aspects.
Also, while we focus on the digital pound and commercial
bank money, the principle of functional consistency
and the analysis framework used in this paper should also be
applicable to emerging forms of private digital money
such as regulated stablecoins \cite{fca-stablecoin-dp}
and to other currencies.}

The contributions of this paper include
\changess{defining and identifying the significance of
functional consistency across all forms of regulated retail
digital money, and 
evaluating the suitability of design options 
to support functional consistency across the digital pound
and commercial bank money.}
We hope the insights presented in this paper will
stimulate discussion, and we look forward to ongoing industry
engagement on retail CBDCs in general and the digital
pound in particular.

%%%%%%%%%%%%%%%%%%%%%%%%%%%%%%%%%%%%%%%%%%%%%%%%%%%%%%%%%%%%%%%%%%%
%% Section 2: The Bank of England Digital Pound Requirements, Design Considerations and Model
%%%%%%%%%%%%%%%%%%%%%%%%%%%%%%%%%%%%%%%%%%%%%%%%%%%%%%%%%%%%%%%%%%%

\section{The Digital Pound}
\label{sec:cbdc-arch-dig-pound-model}

In this section, we summarise the Bank of England's motivations,
key criteria, platform model, potential functional requirements
and design considerations for the digital pound, as described
in the CP and the TWP.

The Bank of England's two primary motivations for the digital pound
are described in the CP \cite{boe-cbdc-cons-paper} \changess{(page 24)}:

\begin{quote}

  \begin{enumerate}

    \item To sustain access to UK central bank money \changess{\textendash} ensuring its
    role as an anchor
    for confidence and safety in our monetary system, and to underpin
    monetary and
    financial stability and sovereignty; and
    
    \item To promote innovation, choice, and efficiency in domestic
    payments as
    our lifestyles and economy become ever more digital.
      
  \end{enumerate}
  
\end{quote}

\noindent
Based on these primary motivations, the Bank of England has
identified the following key criteria for the model of
provision of the digital pound in the CP \cite{boe-cbdc-cons-paper} \changess{(page 51)}:

\begin{quote}

\begin{itemize}

  \item To ensure that central bank money acts as the anchor of monetary and
  financial stability, the model should ensure access to financially risk-free
  central bank money, a direct end-user claim on the Bank [of England]
  and settlement finality for any transactions.

  \item The model should be interoperable with other forms of money, in particular
  cash and bank deposits.

  \item To support innovation, choice and efficiency, the model should be extensible
  and flexible reflecting the fact that the future payments landscape is innovative
  and dynamic.

  \item The model should ensure a standard of operational resilience necessary for
  major national infrastructure. 
    
\end{itemize}

\end{quote}

\begin{figure}[!h]
  \captionsetup{width=15cm}
  \begin{center}
  \includegraphics[width=0.6\linewidth]{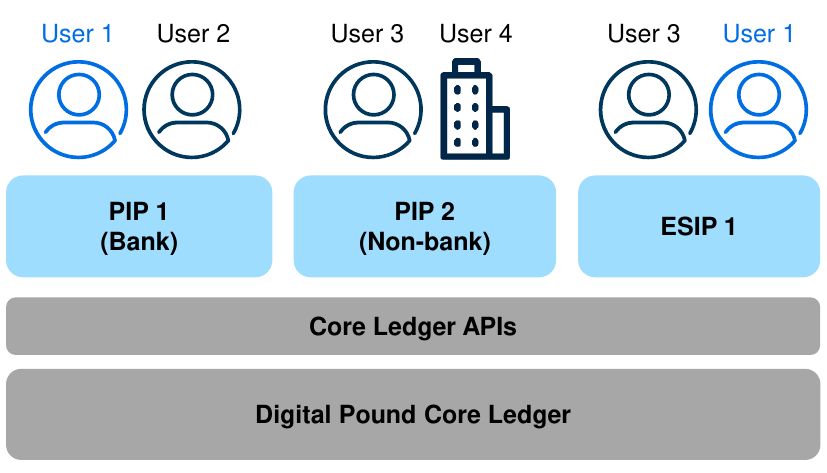}
  \end{center}
  \vspace{-4mm}
  \caption{\footnotesize{
    \changess{The Bank of England's \changess{platform model} for CBDC
    provision, comprising a digital pound
    core ledger, APIs, PIPs, ESIPs and users.}
    Figure adapted from \cite{boe-cbdc-cons-paper}.}}
  \label{fig:uk-cbdc-platmodel-fig}

\end{figure}

\noindent
The platform model 
proposed by the Bank of England,
which we adopt in this paper \changess{(see Figure \ref{fig:uk-cbdc-platmodel-fig})}, comprises the Bank of England operating
a digital pound core ledger and providing access via application programming
interfaces (APIs) 
to authorised and regulated \changes{payment interface providers (PIPs) and
external service interface providers (ESIPs)} that provide
users with access to the digital pound.
The digital pound core ledger would 
record digital pounds issued by the Bank of
England and provide \changess{the} minimum necessary functionality such as
simple payments.
The core ledger APIs would allow PIPs and ESIPs to connect to
the digital pound core ledger and leverage its functionality
to develop and offer innovative services.

%%%%%%%%%%%%%%%%%%%
%% (Start): Figure 1 - The Bank of England Platform Model
%%%%%%%%%%%%%%%%%%%
%%%%%%%%%%%%%%%%%%%
%% (End): Figure 1
%%%%%%%%%%%%%%%%%%%

%\pagebreak 

\noindent
\changess{In the TWP the Bank of England summarised the potential functionality and
features of the digital pound under the following categories
\cite{boe-cbdc-twp} (pages 17-18)}:

\begin{quote}

\begin{itemize}

  \item \emph{Payment devices}: Smart devices and physical cards, existing online
  and in-store point-of-sale infrastructure, Internet of Things devices and wearables.

  \item \emph{Wallet management}: Opening a wallet and viewing balances.
  
  \item \emph{Payments}: Real-time one-off push, peer-to-peer, person-to-business
  (both in-store and online), cross-border, offline, scheduled, micropayments, batch
  payments (e.g. wage) and split payments.

  \item \emph{Interoperability}: Moving between CBDC and other forms of money,
  particularly cash and bank deposits
  
  \item \emph{Economic design}: Ability to implement limits policy.
  
  \item \emph{Identity, data and privacy}: No access to users’ personal data by the Bank
  [of England],
  PIPs undertake know your customer (KYC) checks and anti-money laundering (AML) compliance,
  CBDC would not be anonymous, potential for users to have privacy controls on wallets and
  potential for tiered wallets based on ID information.
    
\end{itemize}

\end{quote}

\noindent
The TWP also describes the design considerations that organise
and guide the Bank of England's work on CBDC technology:

\begin{itemize}

  \item \emph{Privacy}: The CBDC system should be designed to protect user
  privacy, while allowing PIPs and ESIPs the minimum necessary access to
  transaction data needed to provide CBDC services and to fulfil their legal
  and regulatory obligations.

  \item \emph{Security}: The CBDC system should be secure by design to
  ensure confidence in money and promote user trust and adoption.
  
  \item \emph{Resilience}: The CBDC system should be resilient to disruption as
  disruptions may have far-reaching consequences for user confidence, data
  integrity and financial stability.

  \item \emph{Performance}: The CBDC system should be able to handle a high
  number of transactions and confirm and settle these transactions as quickly
  as possible.
  
  \item \emph{Extensibility}: The CBDC system should have an extensible
  design, allowing PIPs and ESIPs to implement additional functionality
  without affecting user services.
  
  \item \emph{Energy usage}: The CBDC system should be, at the very least,
  as energy efficient as existing payment infrastructures and designed
  in a way which minimises any impact on the environment.
    
\end{itemize}

%%%%%%%%%%%%%%%%%%%%%%%%%%%%%%%%%%%%%%%%%%%%%%%%%%%%%%%%%%%%%%%%%%%
%% Section 3: Functional consistency
%%%%%%%%%%%%%%%%%%%%%%%%%%%%%%%%%%%%%%%%%%%%%%%%%%%%%%%%%%%%%%%%%%%

\section{Functional Consistency}
\label{sec:func-cons}

In this section, we explore the concept of functional consistency
for regulated retail digital money \changess{denominated}
in a specific national unit of account.
This includes retail commercial bank money, e-money, and
new forms of money (such as retail CBDCs, tokenised
deposits and regulated stablecoins),
but excludes physical currency notes,
wholesale commercial bank money, and wholesale central bank
money.

The Bank of England and HM Treasury defined ``uniformity of money''
as follows in the CP \cite{boe-cbdc-cons-paper}: 
``all forms of money \changess{\textendash} both bank deposits and cash \changess{\textendash} are valued
equally (`at par' or `face value'), denominated in a common
currency (sterling) and interchangeable with each other'' (page 25).
The CP states that ``the stability of the UK economy and monetary
system relies on the uniformity of money'' (page 25).
Andrew Bailey identified ``singleness of money'' \cite{boe-singleness-money}
as one of the important foundations of money and defined it as
follows:
``\dots wherever we hold our money \changess{\textendash} in bank accounts, notes and coins etc \changess{\textendash} we can
be assured that it all has the same value \changess{\textendash} the pound in my bank account equals the
pound in your account. In other words, money is exchangeable
at par value'' (page 3).
Garratt and Shin identified ``singleness of money'' as a cornerstone
of the modern monetary system \cite{bis-singleness-money}.
However, neither uniformity nor the singleness of money, as
defined above,
include certain operational characteristics of money
that are important to users 
\changess{(such as ease of use and compatibility with existing payment
infrastructures)}.

Fungibility means that any unit of a given form of money is substitutable
for another \cite{riksbank-money-properties}, e.g. one \pounds10 note
can be substituted for two \pounds5 notes and all pounds held in
a commercial bank account are substitutable for each other.
However, a different term is needed \changess{to express the substitutability of units}
of one form of money for units of a different form of money.

Recent developments in new forms of digital money, such as retail
CBDCs and stablecoins, aim to provide users with novel
functionalities, such as programmable payments,
that may not be available in existing forms of money.
While these novel functionalities could allow users to access
innovative capabilities (such as conditional payments), there
is a risk that this could lead to fragmentation in retail
deposits and payments, and \changess{could therefore
\begin{itemize}
  \item negatively impact consumers (e.g. through causing
  confusion) by requiring them to split their liquidity across
  various forms of money in order to obtain the unique operational
  characteristics each of them provides, and
  \item make it more difficult for institutions to maintain stability
  and manage risk.
\end{itemize}}

We define functional consistency for money as follows:

\begingroup
  \advance\leftmargini 2em
\begin{quote}

  {\bf
    {\em Functional consistency is the principle that
     different forms of money have the same
     operational characteristics.}
  }
\end{quote}
\endgroup

\noindent
We believe functional consistency is a desirable principle
for the provision of all forms of retail digital money,
including digital pounds and commercial bank money.
Functional consistency could mitigate the risk of fragmentation and
reduce the negative impacts of fragmentation on consumers,
thereby facilitating both consumer and merchant
adoption of new forms of retail digital money
(including the digital pound).
It could also support consumer choice in domestic payments by
providing common operational characteristics across all forms of
retail digital money.
Note \changess{that} we welcome and encourage the development of novel
functionalities \changess{that} are replicable across all forms of
retail digital money.

We now identify common operational characteristics that 
would ensure functional consistency across 
all forms of regulated retail digital money.
As our starting point, we take the list of properties of 
money identified by Hull and Sattath \cite{riksbank-money-properties}.
We suggest \changess{that} the following initial set
of common operational characteristics are required to
achieve functional consistency:

\begin{itemize}

  \item \emph{Acceptability}: \changess{the probability with which a given form of 
  money is accepted in
  exchange for goods and services provided \cite{minfed-acceptability}}.

  \item \emph{Inclusivity}: \changess{
    the degree to which users can access and use a given form of money
    regardless of their demographics, disabilities, lack of access to technology, skills
    and geographic location
  }.
  
  \item \emph{Divisibility and mergeability}: \changess{
    the ability to divide a given form of money into
 smaller denominations (e.g. to facilitate transactions), and the extent to which units
 can be merged together
  } \cite{riksbank-money-properties}.
  
  \item \emph{Ease of use}: the simplicity with which
  transactions can be conducted with a given form of money 
  (adapted from \cite{riksbank-money-properties}).
  
  \item \emph{Settlement time}: the
  time it takes to settle a transaction in a given form of money
  \cite{riksbank-money-properties}.
  
  \item \emph{Transaction cost}: the pecuniary
  cost, in the form of a fee, imposed on the payer and/or payee when
  performing a transaction using a given form of money
  \cite{riksbank-money-properties}.
  
  \item \emph{Reversibility}: the ability provided 
  to a payer and/or payee to reverse a payment transaction conducted
  using a given form of money, under certain circumstances.
  
  \item \emph{Payment programmability}: the ability of a form
  of money to support both automation and streamlining of the
  payments process \changess{\textendash} \changess{for example, by} triggering payments based on specific
  events or predetermined conditions.

  \item \emph{Transferability}: \changess{
    the ability of a given form of money to be transferred from one
 owner to another
  }
  (adapted from \cite{riksbank-money-properties}).
  
  \item \emph{Confidentiality}: the extent
  to which a \changess{given} form of money protects users' personal and \changess{transaction}
  data from unauthorised access.

  \item \emph{Integrity}: \changess{
    the extent to which the balance and transaction data for a given form
    of money is accurate, trustworthy, complete and protected from accidental or
    unauthorised modification.
  }

  \item \emph{Availability}: the ability for users to
  access and use a given form of money regardless of any technical or
  operational failures.

  \item \emph{Identity-based}: \changess{
    the characteristic of requiring users of a given form of money to disclose 
    their identities while conducting transactions
  } (adapted from \cite{riksbank-money-properties}).
  
  \item \emph{Fungibility}: \changess{
    the substitutability of any unit of a given form of money for 
    another \cite{riksbank-money-properties}.
  }

  \item \emph{Interoperability\footnote
  {We use the term interoperability as an operational characteristic
  of a form of money and not as a technical feature describing
  interaction mechanisms between systems (\changess{as it is used}, for example, in
  \cite{bis-gocb-cbdc-foundfeat}).}}: 
  the ability for
  users to exchange units between two forms of money.

  \item \emph{Payment infrastructure compatibility}: the ability to use
  a given form of money with existing and new payment infrastructures such as
  smart devices, point-of-sale (POS) infrastructure, 
  automated teller machines (ATMs), 
  and payment schemes.

  \item \emph{Support for payment types}: the ability of a given form
  of money to support a variety of payment types such as real-time, 
  batch, offline, cross-border, pull, split, and recurring payments.

\end{itemize}

\noindent
Each of the above common operational characteristics can take on
a range of values (e.g. high acceptability or low transaction cost)
but to achieve functional consistency,
all forms of regulated retail digital money
should have similar values for
these common operational characteristics.

Note that we have excluded the following four types of properties from the
common operational characteristics needed to achieve
functional consistency:

\begin{enumerate}[label={(\arabic*)}]

  \item Properties that are not operational characteristics from a user's
  perspective, such as ``throughput'', ``proof of reserves'' and ``resource
  efficiency''.
  
  \item Properties that do not apply to regulated retail
  digital money, such as ``untraceability'', ``tax avoidability'' and
  ``scarcity''.
  
  \item Properties that are intrinsic to regulated retail
  digital money, such as ``digital'', ``portability'' and
  ``price stability''.
  
  \item Properties that are not appropriate for all forms of money,
  such as ``interest bearing'' and ``user holding limits''.
  For example, commercial bank money can be interest bearing but
  it is \changess{unlikely} that digital pounds will be interest bearing.
  Also, digital pounds may initially have user holding limits but
  commercial bank money does not.

\end{enumerate}

\noindent
\changess{Note that, while we believe functional consistency is a
desirable principle for the provision of all forms of
retail digital money, not all forms of existing private
digital money have the same common operational characteristics.
This lack of common operational characteristics
may cause fragmentation and
potentially negatively impact consumers, which illustrates the
importance of functional consistency for any new form of
public digital money.}

%%%%%%%%%%%%%%%%%%%%%%%%%%%%%%%%%%%%%%%%%%%%%%%%%%%%%%%%%%%%%%%%%%%
%% Section 4: Design Options for Functional Consistency
%%%%%%%%%%%%%%%%%%%%%%%%%%%%%%%%%%%%%%%%%%%%%%%%%%%%%%%%%%%%%%%%%%%

\section{Design Options to Support Functional Consistency} 
\label{sec:uk-cbdc-des-opts}

In our previous paper \changess{\cite{barc-cbdc-iia-doi}} we presented an
illustrative industry architecture \changess{that} aims to place the
digital pound and commercial bank money on a similar footing.
We now present design options that could provide
the common operational characteristics that are required
to achieve functional consistency
across the digital pound and commercial bank money.
These design options are based on both existing payment
ecosystem designs and the Bank of England's platform model
(summarised in Section \ref{sec:cbdc-arch-dig-pound-model}).

\vspace{1mm}

%% Subsection: Related works

\subsection{Existing Designs}
\label{subsec:related-work}

In existing payment \changess{ecosystems some of the
common operational characteristics
(such as payment programmability, interoperability
and} payment infrastructure compatibility) 
are delivered using the following design options:

\begin{itemize}

  \item \emph{By the entity operating the ledger}: 
  For example, commercial banks provide users with the ability
  to open accounts and operate them through various channels.

  \item \emph{\changess{By intermediaries providing end-user access}}: 
  For example, in Open Banking\footnote{The UK Open Banking service
  \changess{allows financial data} and services to be shared between banks
  and third-party service providers securely through the use of APIs to
  enable consumers and businesses to move, manage and make more
  of their money \cite{open-banking}.}, 
  payment initiation service
  providers (PISPs) provide payment initiation services to end
  users \changess{with} payment accounts held at other payment service
  providers (PSPs).

  \item \emph{\changess{By partners of the intermediaries providing end-user access}}: 
  For example, in card schemes, fintech firms use the services of partner
  banks for card issuance and transaction processing.
  
  \item \emph{By common technical service providers\footnote
  {A TSP provides technical services such as communication,
  technical \changess{onboarding and information processing/storage 
  to support authorised financial services providers.}
  It does not have a direct relationship with or provide services to
  end users.
  Description adapted from \cite{fca-agency-models}.}} (TSPs):
  For example, in the UK domestic payment schemes, commercial banks use a
  common \changess{TSP} to integrate with the confirmation
  of payee services provided by other commercial banks.
  Another example is the Swift \cite{abt-swift-network} 
  financial messaging system which
  provides a common
  technical service to facilitate communication between financial
  institutions.

  \item \emph{\changes{By financial market infrastructures (FMIs)\footnote
  {An FMI is a multilateral system among participating institutions,
  including the operator of the system, used for 
  clearing, settling or recording payments, securities, derivatives
  or other financial transactions \cite{bis-cpmi-glossary}.}}}: 
  For example, in the United Kingdom, the Faster Payments System (FPS)
  \cite{abt-fps} provides a central infrastructure that is used by participant banks
  to perform clearing and settlement of instant payments.

\end{itemize}

\vspace{1mm}

%% Subsection:  design options

\subsection{Design Options}
\label{subsec:prop-design-opts}

\begin{figure}[h!]
  \begin{center}
  \includegraphics[width=1\linewidth]{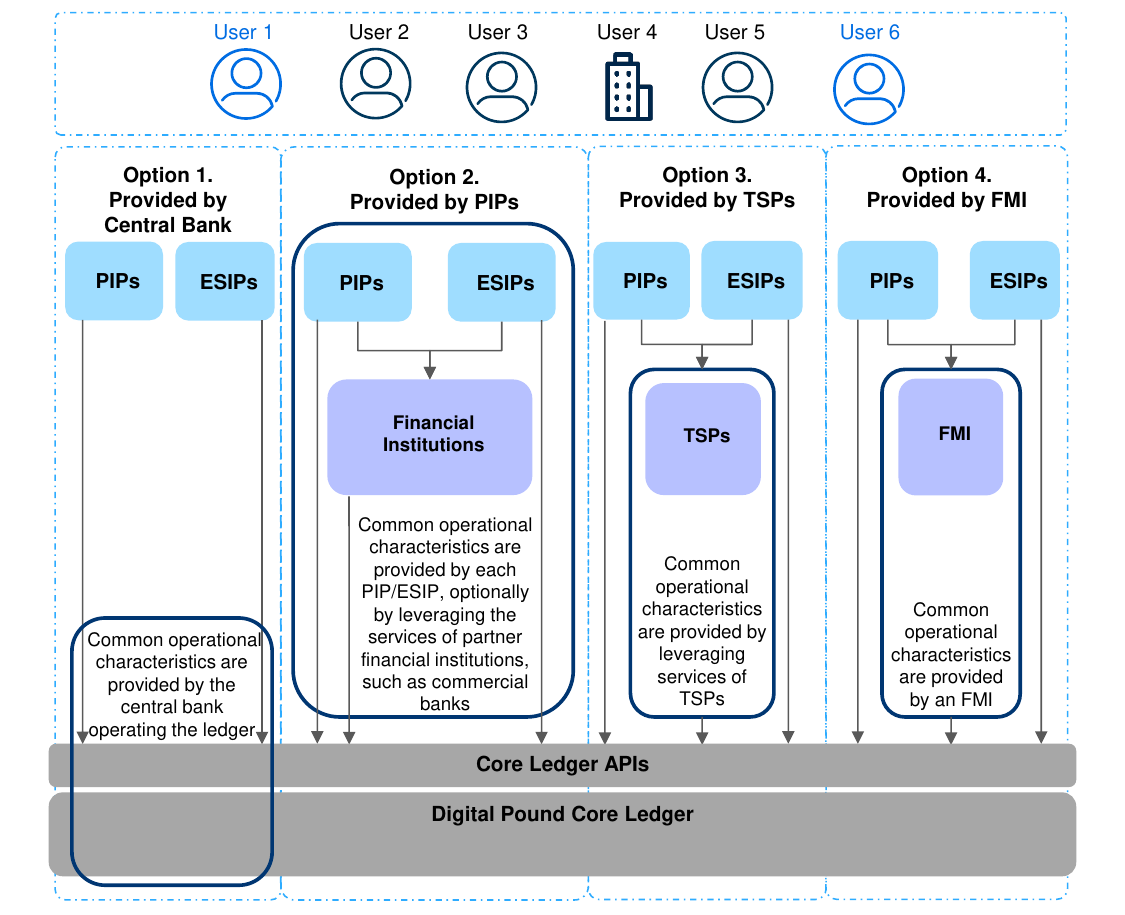}
  \end{center}
  \vspace{-4mm}
  \caption{\footnotesize{Design options that could provide
  the common operational \changess{characteristics required}
  to achieve functional consistency
  across the digital pound and commercial bank money.}}
  
  \label{fig:uk-cbdc-des-opts-fig}

\end{figure}

The design options that 
could provide the common operational characteristics
required
to achieve functional consistency
are presented in Figure
\ref{fig:uk-cbdc-des-opts-fig}.

%%%%%%%%%%%%%%%%%%%
%% (Start): Figure 2 - Design options summary
%%%%%%%%%%%%%%%%%%%
%%%%%%%%%%%%%%%%%%%
%% (End): Figure 2
%%%%%%%%%%%%%%%%%%%

\pagebreak

\vspace{1mm}

\noindent
Note that no single design option could, by itself, provide
functional consistency across all forms of regulated
retail digital money.
In Section \ref{subsec:analysis}, we analyse how a
number of design options could, collectively, provide functional
consistency.
The design options are summarised below:

\begin{itemize}

  \item \emph{Option 1. Provided by central bank}:
    Common operational characteristics are provided by the central bank
    operating the ledger \changess{(e.g. the Bank of England)}.

  \item \emph{Option 2. Provided by PIPs}:
    Common operational characteristics are provided by each PIP/ESIP, optionally by
    leveraging the services of partner financial institutions,
    such as commercial banks.

  \item \emph{Option 3. Provided by TSPs}:
    Common operational characteristics are provided by
    leveraging services of TSPs.

  \item \emph{Option 4. Provided by FMI}:
    Common operational characteristics are provided by an FMI.
    Note that an FMI could also provide technical
    services but we cover such situations under
    option 3 instead.

\end{itemize}

%\vspace{0.25cm}

%%%%%%%%%%%%%%%%%%%%%%%%%%%%%%%%%%%%%%%%%%%%%%%%%%%%%%%%%%%%%%%%%%%
%% Section 5: Architecturally significant use cases and capabilities
%%%%%%%%%%%%%%%%%%%%%%%%%%%%%%%%%%%%%%%%%%%%%%%%%%%%%%%%%%%%%%%%%%%
\section{Architecturally Significant Use Cases and Key Capabilities}
\label{sec:uk-cbdc-uc-cap}

\changess{To evaluate} the suitability of the design
options, we now present some architecturally significant use cases
and the key capabilities that support them.
Note that \changess{
we focus on the digital pound and commercial bank
money in our use cases, and
we do not elaborate some of the use cases for commercial
bank money (such as opening an account) as they have
established solutions.}

%% Subsection: User personas

\subsection{User Personas, Use Cases and Supporting Capabilities}
\label{subsec:user-personas-cases}

\changess{We identify for the digital pound two user personas} with differing
demographics, abilities and motivations:

\begin{itemize}

  \item \emph{Laura}: Laura is a 20-year-old college student
  and a digital native.
  She primarily uses her smartphone and her laptop for all her
  financial needs, with
  occasional use of ATMs for a few cash transactions.
  She has an existing commercial
  bank account.

  \item \emph{Bob}: Bob is a 70-year-old pensioner who relies on
  his bank branch and
  ATMs for his financial needs. 
  He is not comfortable with using smartphones
  and websites for conducting financial transactions due to a
  lack of internet access.

\end{itemize}

\noindent
Laura's and Bob's customer journeys are presented in Table
\ref{table:table-uc-cap}, with the architecturally-significant
use cases and capabilities that support them.
Note \changess{that} some of the use cases have been adapted from challenges
posed in the Barclays CBDC Hackathon 2022 \cite{barclays-cbdchack-challenges} and from
Barclays' prototype
of a novel use case of ``payment on physical delivery'' developed
as part of Project Rosalind 
\cite{bisihl-rosalind-report}.

\pagebreak
%%%%%%%%%%%%%%%%%%%
%% (Start): Longtable - Use cases and capabilities
%%%%%%%%%%%%%%%%%%%

\begin{center}
  \begingroup
  \renewcommand{\arraystretch}{1.5}     % Default 1
  \captionsetup{width=15cm}
    
    \begin{longtable}{ >{\raggedright}p{0.3\textwidth} >{\arraybackslash}p{0.6\textwidth} }
     
     \toprule
     \textbf{Use Case} & \textbf{Supporting Capabilities} \\  [0.5ex] 
     \midrule
     \endfirsthead

     \multicolumn{2}{r}{\textit{Continued from previous page}} \\
     \toprule
     \textbf{Use Case} & \textbf{Supporting Capabilities} \\  [0.5ex] 
     \midrule
     \endhead

     \bottomrule
     \multicolumn{2}{r}{\textit{Continued on next page}} \\
     \endfoot

     \bottomrule
     \caption{Customer journeys of two user personas 
       and the architecturally-significant
       use cases and capabilities that support the user journeys.}
     \label{table:table-uc-cap}
     \endlastfoot

      % Table body  

      % Laura's consumer journey
        
      \multicolumn{2}{l}{\textit{(a) Laura's customer journey}} \\
      
      (1) Create new digital pound wallet using a mobile app
      provided by a PIP   &
      \begin{compactimize}
        \item \changess{Completion of KYC checks}
        \item Wallet creation (create, set authentication etc)
        \item Alias management (selection/assignment of alias)
        \item \changess{Setup of user holding limits across PIPs}
        \vspace{0.5cm}
      \end{compactimize} \\

      (2) Fund a digital pound wallet from a commercial bank account
      using the app  &
      \begin{compactimize}
        \item \changess{Integration} of the app with Open Banking to initiate
          funding payments (e.g. using FPS)
        \item Interoperability between commercial bank money
          and digital pounds, including transfer of confidential
          payment information
        \item \changess{Enforcement of} user holding limits
        \vspace{0.5cm}
      \end{compactimize} \\

      (3) Order a product from online \changess{merchant, and on delivery,
       make a payment to merchant's commercial bank account using a digital pound alias} &
      \begin{compactimize}
        \item \changess{Integration of} merchant payment gateways with the digital
          pound ecosystem
        \item Digital pound alias validation (initiated by merchant
          or payment gateway)
        \item Programmable payments using digital pounds 
          (lock funds at purchase and pay on delivery)
        \item Interoperability between digital pounds and 
          commercial bank money (e.g. using FPS), 
          including transfer of confidential payment information
          between the merchant's payment gateway and Laura's PIP
          \vspace{0.5cm}
      \end{compactimize} \\

      (4) Order a product from online merchant and pay on delivery from
      Laura's commercial bank account to merchant's
      commercial bank account &
      \begin{compactimize}
        \item Merchant payment gateway integration with Open Banking
        \item Programmable payments using commercial bank money
          (lock funds at purchase and pay on delivery)
        \item \changess{Transferral of funds} between commercial bank
          accounts (e.g. using FPS),
          including transfer of \linebreak confidential payment information
          between the \linebreak merchant's payment gateway and Laura's bank
          \vspace{0.5cm}
      \end{compactimize} \\ 

      \midrule

      % Bob's customer journey

      \multicolumn{2}{l}{\textit{(b) Bob's customer journey}} \\
       
      (1) \changess{Create a} new digital pound wallet at a point-of-presence (e.g. post
      office) and collect a physical smart card  &
      \begin{compactimize}
        \item \changess{Completion of} KYC checks
        \item Wallet creation (create, set authentication, etc)
        \item \changess{Integration} with card schemes (smart card issuance)
        \item Alias management (linking card number to digital pound wallet)
        \item \changess{Setup of }user holding limits across PIPs
        \vspace{0.5cm}
      \end{compactimize} \\

      (2) Deposit physical cash in a digital pound wallet at a
      point-of-presence (manual/automated) using the
      smart card  &
      \begin{compactimize}
        \item \changess{Integration} with card schemes (card validation 
          and user authentication)
        \item Physical cash acceptance and verification
        \item \changess{Integration} with card schemes (transaction clearing
           and settlement for payment from the point-of-presence operator's
           commercial bank account to Bob's digital pound wallet,
           including transfer of confidential payment information)
        \item \changess{Enforcement of} user holding limits
        \vspace{0.5cm}
      \end{compactimize} \\

      (3) Buy \changess{a} product from local store and pay to merchant's
       digital pound wallet using the smart card (linked to Bob's
       digital pound wallet) &
      \begin{compactimize}
        \item \changess{Integration} with card schemes (card validation, 
          user authentication and debit authorisation)
        \item \changess{Integration} with card schemes (transaction clearing
          and settlement for payment from one digital pound wallet to 
          another, including transfer of confidential payment
          information)
        \item \changess{Enforcement of} holding limits for merchants, if applicable
        \vspace{0.5cm}
      \end{compactimize} \\

      (4) Withdraw physical cash at an ATM using the smart card
      (linked to Bob's digital pound wallet) \changess{in order to pay} in cash at a
      farmer's market  &
      \begin{compactimize}
        \item \changess{Integration} with ATM network schemes (card \newline
        validation, user authentication and debit authorisation)
        \item \changess{Integration} with ATM network schemes (transaction \newline
          clearing and settlement for payment from Bob's digital
          pound wallet to the ATM operator's commercial bank account,
          including transfer of confidential payment information)
        \item Physical cash disbursement
        \vspace{0.5cm}
      \end{compactimize} \\ 

    \end{longtable}

  \endgroup
      
\end{center}
      
%%%%%%%%%%%%%%%%%%%
%% (End): Long Table - Use cases and capabilities
%%%%%%%%%%%%%%%%%%%

%% Subsection: Key Capabilities

\subsection{Key Capabilities}
\label{subsec:key-capabilities}

We now select a subset of key capabilities 
(from the supporting capabilities presented in Table \ref{table:table-uc-cap})
that will be used to evaluate the suitability of the design options
(presented in Section \ref{subsec:prop-design-opts}).
We exclude setting and enforcing user holding limits because these
capabilities support operational characteristics which are not
appropriate for all forms of money \changess{and that are therefore}
not required for functional consistency.
We also exclude other supporting capabilities (such as wallet
creation, integrating apps with Open Banking,
funds transfer between commercial bank accounts, smart card issuance, and
physical cash acceptance and disbursement)
because they have established solutions.

These key capabilities support some of the common operational
characteristics (presented in Section \ref{sec:func-cons}) that
are needed to achieve functional consistency,
such as
acceptability, inclusivity, ease of use, reversibility,
payment programmability, transferability, confidentiality,
interoperability, payment infrastructure compatibility and support
for various payment types.
Table \ref{table:table-key-cap} lists the key capabilities selected,
with their motivations and the common operational characteristics they
support.

%%%%%%%%%%%%%%%%%%%
%% (Start): Longtable - Key capabilities
%%%%%%%%%%%%%%%%%%%

\begin{center}
  \begingroup
  \renewcommand{\arraystretch}{1.5}     % Default 1
  \captionsetup{width=15cm}
    
    \begin{longtable}{ >{\raggedright}p{0.3\textwidth} >{\raggedright}p{0.35\textwidth} >{\arraybackslash}p{0.28\textwidth} }
     
     \toprule
     \textbf{Motivation} & \textbf{Key Capability} & \textbf{Operational Characteristics} \\  [0.5ex] 
     \midrule
     \endfirsthead

     \multicolumn{3}{r}{\textit{Continued from previous page}} \\
     \toprule
     \textbf{Motivation} & \textbf{Key Capability} & \textbf{Operational Characteristics} \\  [0.5ex] 
     \midrule
     \endhead

     \bottomrule
     \multicolumn{3}{r}{\textit{Continued on next page}} \\
     \endfoot

     \bottomrule
     \caption{Key capabilities with their motivations, and the common
        operational characteristics they support.}
     \label{table:table-key-cap}
     \endlastfoot

      % Table body

      % C1.

      The digital pound model described in the CP states that
      PIPs/ESIPs would be responsible for recording the identity
      of digital pound users and carrying out necessary 
      KYC checks.    & 

      \textit{C1. Completing necessary KYC checks}.\newline
      The capability for PIPs/ESIPs to perform KYC checks, either
      by themselves or by leveraging certified digital identity
      service providers
      e.g. OneID, Onfido and Experian \cite{uk-idsp-list}. &

      \begin{compactimize}
        \item Supports \emph{identity-based} access
      \end{compactimize} \\ 

      \midrule
      % C2.

      The digital pound model described in the CP states that
      the Bank of England, as operator of the CBDC system, would
      not have access to personal data of users \changess{
      (note that most retail payment instructions contain both
      personally identifiable information (PII),
      such as names and addresses\footnote
      {\changess{Article 4(1) of the UK General Data Protection
      Regulation 2016 \cite{uk-gdpr} includes names,
      identification numbers and location data as examples of
      personal data.
      Of these, the names and addresses of payers and payees
      are mandatory fields in some payment message formats such as
      the CHAPS pacs.008.001.08 (Single Customer Credit Transfer)
      format \cite{boe-chaps-pacs008}.}},
      and confidential information, such as purpose of
      payment).}     & 

      \textit{C2. \changess{Transferring confidential} payment information
        between PIPs}.\newline
      The capability to transfer PII and confidential information
      for a digital pound payment 
      from the payer to the payee, without it being accessible to the
      Bank of England.  &

      \begin{compactimize}
        \item Protects user \newline \emph{confidentiality}
      \end{compactimize} \\ 
      
      \midrule
      % C3.

      The TWP envisages the use of a range of different aliases
      that could be used to route transactions between users, 
      conceal identifiers of digital pound wallets on the core ledger
      and enable interoperability with existing payment
      infrastructures. & 

      \textit{C3. Alias management for digital pound wallets}.\newline
      The capability to generate, store, lookup and validate aliases
      for digital pound wallets,
      including user-selected aliases (e.g. mobile numbers) and 
      assigned aliases (e.g. card numbers). &

      \begin{compactimize}
        \item Protects user \newline \emph{confidentiality}
        \item Supports \emph{ease of use} for payment initiation
        \item Supports \emph{interoperability} with existing payments systems 
      \end{compactimize} \\ 

      % C4.

      The CP identifies interoperability with bank deposits
      as one of the key criteria for the digital pound model.    & 

      \textit{C4. Interoperability with commercial bank money}.\newline
      The capability to transfer value between digital
      pound wallets and commercial bank accounts,
      including the transfer of confidential payment information and
      settlement finality.  &

      \begin{compactimize}
        \item Improves \emph{acceptability} \newline with people/merchants who
          \changess{do not} have digital pound wallets
        \item Supports \emph{transferability} and 
           \emph{interoperability}
        \item Supports \emph{payment infrastructure compatibility}
      \end{compactimize} \\ 

      \midrule
      % C5.

      The CP states that programmability could support
      innovation through improved functionality for users.  & 

      \textit{C5. Programmable payments}.\newline
      The capability to both automate and streamline the payments
      process \changess{\textendash} \changess{for example, by triggering payments based on specific events
      or predetermined conditions (e.g. users can lock funds at purchase and drawdown the
      locked funds and pay the merchant on delivery)}.  &

      \begin{compactimize}
        \item Supports \emph{payment \newline programmability} across all forms
          of money
        \item Improves \emph{ease of use} \newline and \emph{integrity}
          by \newline automating payments
        \item Supports recurring and split \emph{payment types}
        \item Supports automated \newline rules-based \emph{reversibility}
      \end{compactimize} \\ 

      \midrule
      % C6.

      The TWP lists person-to-business payments
      using existing online and in-store POS
      infrastructure as potential functionality of the digital
      pound.  &

      \textit{C6. Integrating merchant payment gateways}.\newline
      The capability to integrate payment gateways
      used by merchants with the digital pound ecosystem.
      This includes initiating payments using digital pound aliases,
      validating aliases, requesting payment, and settling funds
      between digital pound wallets and commercial bank \changess{accounts}
      (including transfer of confidential payment information). &

      \begin{compactimize}
        \item Improves \emph{ease of use} \newline and
         \emph{acceptability} at \newline merchants
        \item Supports \emph{transferability} and \emph{interoperability}
        \item Supports \emph{payment infrastructure compatibility}
      \end{compactimize} \\ 

      %\midrule
      % C7.

      The TWP identifies smart cards as potential
      payment devices for the digital pound that can
      be used at existing online and in-store POS
      infrastructure\changess{; not all existing POS infrastructure
      may be upgraded to integrate directly with the digital
      pound ecosystem, so payments via card schemes would
      need to be supported}. &

      \textit{C7. Integrating with card schemes}.\newline
      The capability to integrate with card schemes
      (e.g. Visa \cite{abt-visa} and Mastercard \cite{abt-mastercard})
      to perform smart card validation,
      user authentication, debit authorisation, clearing and
      settlement across digital pounds and commercial bank money
      (including transfer of confidential payment information).  &

      \begin{compactimize}
        \item Improves \emph{ease of use} \newline and
         \emph{acceptability} at \newline merchants
        \item Improves \emph{inclusivity} \newline using physical smart \newline cards
        \item Leverages card schemes for \emph{reversibility}, \newline 
          \emph{transferability} and \newline \emph{interoperability}
        \item Supports \emph{payment infrastructure compatibility} and \newline
          various \emph{payment types}
          \vspace{0.35cm}
      \end{compactimize} \\ 

      \midrule
      % C8.

      The CP identifies interoperability between digital pounds and
      physical cash as a key feature of a digital pound model\changess{
      ; not all existing ATMs
      may be upgraded to integrate directly with the digital
      pound ecosystem, so transactions via ATM network schemes would
      need to be supported.} &

      \textit{C8. Integrating with ATM network schemes}.\newline
      The capability to integrate with ATM
      network schemes (e.g. LINK \cite{abt-link-network}) to
      support cash withdrawal, cash deposit and, potentially,
      other features such as balance \changess{checks}, transfers and 
      mini statements
      (including transfer of confidential payment information).  &

      \begin{compactimize}
        \item Supports \emph{interoperability} with \newline physical cash
        \item Improves \emph{inclusivity} and \emph{ease of use} for
          users who need physical cash
          \item Supports \emph{payment infrastructure compatibility}
      \end{compactimize} \\ 

    \end{longtable}

  \endgroup
      
\end{center}

%%%%%%%%%%%%%%%%%%%
%% (End): Long Table - Key capabilities
%%%%%%%%%%%%%%%%%%%

%%%%%%%%%%%%%%%%%%%%%%%%%%%%%%%%%%%%%%%%%%%%%%%%%%%%%%%%%%%%%%%%%%%
%% Section 6: Evaluation of Design Options
%%%%%%%%%%%%%%%%%%%%%%%%%%%%%%%%%%%%%%%%%%%%%%%%%%%%%%%%%%%%%%%%%%%

\section{Evaluation of Design Options} 
\label{sec:uk-cbdc-des-opts-anal}

We now evaluate the suitability of the design options
(presented in Section \ref{subsec:prop-design-opts}) to provide
the key capabilities (identified in Section \ref{subsec:key-capabilities}).
We describe the assumptions that underpin our evaluation, 
define a suitability rating model, 
evaluate the design options against the key capabilities, and 
draw some initial insights.

%% Subsection: Assumptions

\subsection{Assumptions}
\label{subsec:assumptions}

The following assumptions \changess{are} made while evaluating the
design options:

%\pagebreak

\begin{itemize}

  \item \emph{Alias management}:
  
  \begin{itemize}  
    \item Any given alias (such as a user's mobile number) would be
    mapped to a single digital pound wallet and not reused
    across multiple wallets.
    \item Alias \changess{look-up} and validation services would be accessible
    to PIPs, ESIPs and
    other participants of the digital pound ecosystem.
  \end{itemize}

  \item \emph{Interoperability between digital pounds and commercial
    bank money}: 
    
    \begin{itemize}
      \item \changess{The FPS}\footnote
        {The future UK New Payments Architecture (NPA) \cite{payuk-npa}
        may provide interoperability between digital pounds and
        commercial bank money.}
        would be used to transfer funds between digital
        pound wallets and commercial bank accounts, as it provides
        immediate clearing for \changess{low-value} payments.
        \changess{However, the FPS would only use settlement accounts at the Bank}
        of England to perform settlement and would not provide
        settlement directly on the digital pound core
        ledger.
      \item The Bank of England would integrate the new digital
        pound core ledger with \changess{its} existing reserve/settlement
        account ledger to enable fund transfers between them.
        However, the digital pound core ledger will not be
        directly integrated with commercial bank ledgers for
        funding or defunding.
    \end{itemize}

  \item \emph{Issuing smart cards linked to digital pond
    wallets}:

    \begin{itemize}

      \item The Bank of England would
      not get involved in issuing smart cards to digital
      pound users (e.g. assigning common Bank Identification Number\footnote
      {\changess{Bank identification numbers are the first six digits of a bank card number or payment card number
      (specified in standard ISO/IEC 7812 of the International Organization for Standardization and
      International Electrotechnical Commission) used in credit cards, debit cards, stored-value cards,
      etc, to identify the card brand, the issuing institution or bank, the country of issuance, the card type
      and the category of the card.}}
             
    \end{itemize}

  \item \emph{Support for existing payment infrastructures}:

    \begin{itemize}
      
      \item Merchants would use existing online and in-store POS
      infrastructure to accept digital pound payments from users.
      However, not all POS infrastructure would be upgraded to
      integrate directly
      with the digital pound ecosystem\changess{, so} existing means
      of payment (e.g. using cards or Open Banking) would need to
      be supported.
      
      \item Users would be able to withdraw cash from their
      digital pound wallets using existing ATMs.
      However, not all ATMs would be upgraded to integrate directly
      with the digital pound ecosystem\changess{, so} existing means
      of operation (e.g. using cards and ATM network schemes)
      would need to be supported.
      
      \item Merchant acquirers\footnote
      {Merchants partner with financial institutions
      that provide payment gateway, payment processor and merchant
      \changess{accounts}, thereby allowing end-to-end processing and settlement of
      card transactions.
      In this paper, we refer to these financial institutions
      as merchant acquirers.} 
      would be participants of UK payment systems.
      However, not all merchant acquirers would
      be PIPs because some merchant acquirers may not wish to take
      on all the responsibilities of a PIP.
      
    \end{itemize}

\end{itemize}

\vspace{1mm}
%% Subsection: Suitability ratings

\subsection{Suitability Ratings}
\label{subsec:suitability-ratings}

We now define suitability ratings
(``suitable'', ``partially suitable'' or ``unsuitable'')
\changess{that} will subsequently be used to assess each design option
against each key capability:

\begin{itemize}

  \item \emph{Suitable}:
    The design option could provide the key capability in alignment
    with the Bank of England's key criteria, requirements and technology
    design considerations.

  \item \emph{Partially suitable}:
    The design option could provide the key capability either in 
    partial alignment with the Bank of England's key criteria,
    requirements and technology design considerations, 
    or by introducing additional complexity, risk or cost
    in comparison \changess{with} other design options.

  \item \emph{Unsuitable}:
    The design option could provide the key capability but in
    conflict with the Bank of England's key criteria,
    requirements and technology design considerations,
    or
    by introducing substantial additional complexity, risk or
    cost
    in comparison \changess{with} other design options.

\end{itemize}

%\vspace{1mm}

%% Subsection: Analysis

\subsection{Analysis}
\label{subsec:analysis}

\begin{figure}[!h]
  \captionsetup{width=15cm}
  \begin{center}
  \includegraphics[width=1\linewidth]{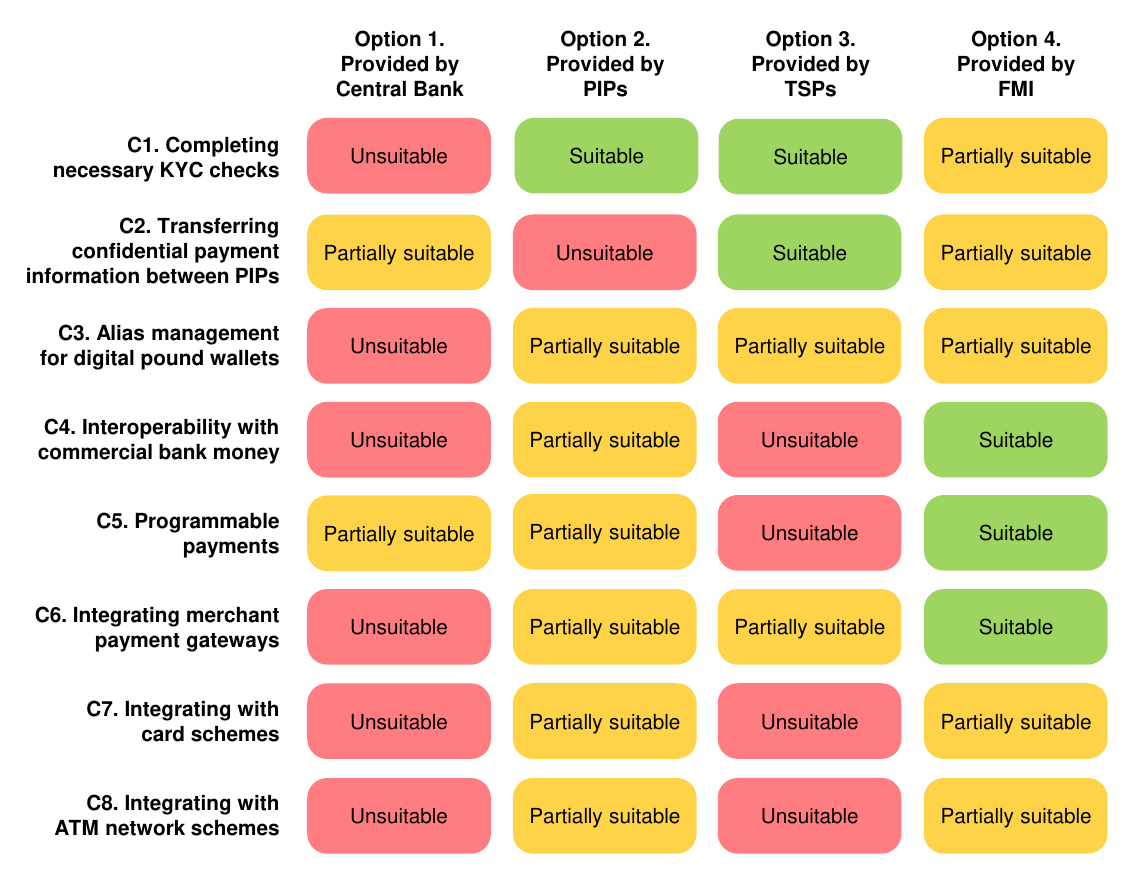}
  \end{center}
  \vspace{-4mm}
  \caption{\footnotesize{\changess{
    Summary of preliminary evaluation of each design option's suitability to 
    provide each key capability.
    These design options are for provision of key capabilities by
    the central bank, PIPs,
    TSPs or 
    an FMI.}
    Note that, for some key capabilities, there is no single
    design option that is suitable and, instead, a combination of
    design options may be suitable.}}
  \label{fig:uk-cbdc-cap-heatmap}

\end{figure}

This section includes 
a summary of our preliminary evaluation of each design option's 
suitability 
\changess{for providing each key capability
(in Figure \ref{fig:uk-cbdc-cap-heatmap}), the preliminary evaluation itself (in
Table \ref{table:tab-des-opts-analysis}),
and some initial insights.}

%%%%%%%%%%%%%%%%%%%
%% (Start): Figure 3 - Capability heatmap
%%%%%%%%%%%%%%%%%%%
%%%%%%%%%%%%%%%%%%%
%% (End): Figure 3
%%%%%%%%%%%%%%%%%%%

\vspace{1mm}
%%%%%%%%%%%%%%%%%%%
%% (Start): Longtable - Design options analysis
%%%%%%%%%%%%%%%%%%%

\begin{landscape}
\begin{center}
  \begingroup
  \renewcommand{\arraystretch}{1.2}     % Default 1
  \captionsetup{width=20cm}
    
    \begin{longtable}{ >{\raggedright}p{0.17\textwidth} >{\raggedright}p{0.33\textwidth} >{\raggedright}p{0.31\textwidth} >{\raggedright}p{0.3\textwidth} >{\raggedright\arraybackslash}p{0.31\textwidth} }
     
     \toprule
       \textbf{Capability} & 
       \textbf{Option 1. \linebreak Provided by Central Bank} & 
       \textbf{Option 2. \linebreak Provided by PIPs} & 
       \textbf{Option 3. \linebreak Provided by TSPs} & 
       \textbf{Option 4. \linebreak Provided by FMI} \\  [0.5ex] 
     \midrule
     \endfirsthead

     \multicolumn{5}{r}{\textit{Continued from previous page}} \\
     \toprule
      \textbf{Capability} & 
      \textbf{Option 1. \linebreak Provided by Central Bank} & 
      \textbf{Option 2. \linebreak Provided by PIPs} & 
      \textbf{Option 3. \linebreak Provided by TSPs} & 
      \textbf{Option 4. \linebreak Provided by FMI} \\  [0.5ex] 
     \midrule
     \endhead

     \bottomrule
     \multicolumn{5}{r}{\textit{Continued on next page}} \\
     \endfoot

     \bottomrule
     \caption{Preliminary evaluation of each design option's suitability to
      provide each key capability.}
     \label{table:tab-des-opts-analysis}
     \endlastfoot

      % Table body:

      C1. Completing necessary KYC checks &

      \emph{Unsuitable} \linebreak 
      KYC capabilities (such as integrating with
      identity providers) could be provided 
      \changess{to PIPs/ESIPs by the CBDC system
      operated by the Bank of England.}
      However, this
      could result in users' PII
      being available to the Bank of England. &
      
      \emph{Suitable} \linebreak 
      Each PIP/ESIP could either perform KYC checks by
      \changess{itself} or integrate with one or more digital identity
      service providers. &

      \emph{Suitable} \linebreak 
      Each PIP/ESIP could either perform KYC checks by
      \changess{itself or integrate with one or more digital identity
      service providers.
      In addition,} TSPs could simplify the onboarding and
      integration of PIPs/ESIPs
      with multiple digital identity service providers. &

      \emph{Partially suitable} \linebreak 
      Supporting KYC checks and
      integrating PIPs/ESIPs with identity service providers
      falls outside the typical scope of an FMI. \\

      \midrule

      C2. \changess{Transferring confidential} payment information between PIPs &

      \emph{Partially suitable} \linebreak 
      Core ledger APIs could support encrypted payment
      information that is only visible to payers, payees and their
      respective PIPs, \changess{including APIs for secure distribution of
      encryption keys between PIPs.}
      However, if keys are compromised, confidential
      payment information could become accessible to the central bank. &

      \emph{Unsuitable} \linebreak 
      Each PIP could onboard and integrate with all other PIPs in
      order to transfer payment information on a separate peer-to-peer
      channel\changess{;
      this} requires all existing PIPs to onboard and integrate with
      every new PIP. &

      \emph{Suitable} \linebreak 
      TSPs could simplify the onboarding and integration of PIPs
      and enable PIPs to share confidential payment information
      on a separate peer-to-peer channel. &
      
      \emph{Partially suitable} \linebreak 
      Digital pound payments \changess{would be} cleared and settled
      using the core ledger APIs\changess{;
      transferring} confidential payment information between PIPs
      for such payments falls outside the typical scope
      of an FMI. \\

      C3. Alias management for digital pound wallets &

      \emph{Unsuitable}\footnote
      {It is currently unclear whether
      novel privacy
      enhancing technologies could allow
      a CBDC system operated by a central bank,
      as per the platform model, to support
      alias management without a risk of compromising user privacy
      in future.}
      \linebreak 
      Aliases could be stored in the CBDC system operated by the Bank
      of England using a protection method (e.g. a one-way hash
      function).
      However, this could allow the Bank of England to
      identify the user of a
      digital pound wallet from a known alias (e.g. \changess{a} mobile number).  &

      \emph{Partially suitable} \linebreak 
      Each PIP could manage and store the aliases (which include a PIP 
      identifier) of its users, and give access to other PIPs
      to lookup and validate aliases on a separate peer-to-peer channel\changess{; this} requires all existing PIPs to onboard and integrate with
      every new PIP. &
      
      \emph{Partially suitable} \linebreak 
      Each PIP could manage and store the aliases (which include a PIP 
      identifier) of its users, and give access to other PIPs
      to lookup and validate aliases on a separate peer-to-peer channel.
      TSPs would simplify the registration and integration of PIPs. &
      
      \emph{Partially suitable} \linebreak 
      PIPs could \changess{use} an FMI service to protect aliases 
      (e.g. one-way hashing function or mapping table) and,
      potentially, to map protected aliases to digital
      pound wallets.
      However, providing this service falls outside the
      typical scope of an FMI. \\

      \midrule

      C4. \linebreak Interoperability with commercial bank money &

      \emph{Unsuitable} \linebreak 
      Integrating the CBDC system operated by the Bank of England
      directly with existing payments systems could require the
      CBDC system to process PII (\changess{e.g.} payer and
      payee details) contained in the message formats of most
      payment systems.\footnote
      {If the CBDC system \changess{integrated directly
      with payment systems (without processing PII), the} 
      payment systems would need to be enhanced to support
      encrypted fields in messages and 
      a participant key exchange mechanism that supports field
      encryption.} &
      
      \emph{Partially suitable} \linebreak 
      In order to settle payments between digital
      pounds and commercial bank money,
      each PIP could either be a direct participant of UK payment
      systems or an indirect participant via a financial institution
      that is also a PIP\changess{;}
      PIPs that are not existing participants may incur 
      higher costs. &
      
      \emph{Unsuitable} \linebreak 
      TSPs would not be direct participants of UK payment systems.
      Each PIP would integrate with its direct participant \changess{in} the UK
      payment systems\changess{; introducing} a TSP in this integration would
      increase complexity. &
      
      \emph{Suitable} \linebreak 
      An FMI could be a participant of the UK payments systems and
      hold a technical settlement account on both 
      the Bank of England settlement ledger and the digital pound
      core ledger\changess{; the} FMI could then clear and settle payments between
      digital pounds and commercial bank money. \\

      C5. Programmable payments\footnote{Some programmability
        use cases 
        (e.g. the ``pay on delivery'' use cases in Table 1)
        require locking functionality to earmark funds.
        In this paper, \changess{we assume locking functionality
        would be provided for each form of money} by either the
        ledger systems via APIs or by the programmability layer.} &

      \emph{Partially suitable} \linebreak 
      A full programmability infrastructure in the CBDC
      system
      would not align with the platform model
      principle that the core ledger should
      provide the minimum necessary functionality. 
      Also, this may result in 
      functionally inconsistent programmability 
      across all forms of money. &
      
      \emph{Partially suitable} \linebreak 
      A separate programmability infrastructure at each PIP would
      require each PIP to bear the cost and complexity
      of operating the infrastructure.
      Also, this may result in
      functionally inconsistent programmability
      across PIPs and across other forms of money. &
      
      \emph{Unsuitable} \linebreak 
      Providing a programmability layer falls outside the
      scope of TSPs as it involves automating payments
      based on events and conditions and, potentially,
      integrating with payment systems. &
      
      \emph{Suitable} \linebreak 
      An FMI could provide a common programmability layer that
      integrates with all relevant ledgers (such as the
      digital pound core ledger and
      commercial \changess{banks'} ledgers) and payment systems, thereby
      enabling programmability that is functionally consistent
      across all forms of money. \\

      \midrule

      C6. Integrating merchant payment gateways &

      \emph{Unsuitable} \linebreak 
      Merchant acquirers \changess{would} need access
      to the core ledger APIs for payment processing,
      but not all of them would be PIPs/ESIPs.
      \changess{In addition}, this option is unsuitable for
      alias validation and interoperability with commercial
      bank money 
      (see rows C3 and C4). &

      \emph{Partially suitable} \linebreak 
      Merchant acquirers would need to integrate with every
      PIP in order to validate aliases and process payments.
      However, this option is not fully suitable for
      alias validation and interoperability with
      commercial bank money 
      (see rows C3 and C4). &     
      
      \emph{Partially suitable} \linebreak 
      TSPs could simplify the integration of merchant
      acquirers with PIPs in order to validate aliases and
      process payments.
      However, this option is unsuitable for
      interoperability with commercial bank money.
      (see row C4). &
      
      \emph{Suitable} \linebreak 
      An FMI could simplify the integration of merchant
      acquirers with PIPs in order to validate aliases and
      process payments.
      \changess{In addition}, this option is suitable for
      interoperability with commercial bank money 
      (see rows C3 and C4).
      \\

      C7. Integrating with card schemes &

      \emph{Unsuitable} \linebreak 
      Integrating the CBDC system operated by the Bank of England
      directly with card schemes could require the CBDC system
      to process PII (\changess{e.g.} card
      numbers and card holder names) contained in card
      scheme messages.\footnote
      {If the CBDC system \changess{integrated directly
      with card schemes (without processing PII),} the 
      card schemes would need to be enhanced to support
      encrypted fields in messages and 
      a member key exchange mechanism that supports field
      encryption.} &
      
      \emph{Partially suitable} \linebreak 
      PIPs would either be principal issuing members of card
      schemes or partner with such scheme members\changess{; they}
      would use core ledger APIs to process transactions
      and would
      interoperate with commercial bank money for settlement
      (see row C4). &
      
      \emph{Unsuitable} \linebreak 
      TSPs would not be card scheme members and
      are unsuitable for providing interoperability
      with commercial bank money (see row C4).
      Also, introducing a TSP between a PIP and its principal
      issuing member would increase complexity. &
      
      \emph{Partially suitable} \linebreak 
      Becoming a card scheme member falls outside the typical
      scope of an FMI.
      An FMI could support card transaction settlements between
      digital pounds and commercial bank money
      (see row C4),
      with minimal change to card schemes.  \\

      \midrule

      C8. Integrating with ATM network schemes &

      \emph{Unsuitable} \linebreak 
      Integrating the CBDC system operated by the Bank of England
      directly with ATM network schemes would require the CBDC system to
      process PII (\changess{e.g.} card
      numbers and card holder names)
      contained in ATM network scheme messages. &
      
      \emph{Partially suitable} \linebreak 
      PIPs would either be members of ATM network schemes, or 
      partner with existing members\changess{; they} would use core ledger APIs to process transactions
      and would
      interoperate with commercial bank money for settlement
      (see row C4). &

      \emph{Unsuitable} \linebreak 
      TSPs would not be ATM network scheme members and
      are unsuitable for providing interoperability
      with commercial bank money (see row C4).
      Also, introducing a TSP between a PIP and its ATM
      network partner would
      increase complexity. &
      
      \emph{Partially suitable} \linebreak 
      Becoming an ATM network scheme member falls outside the typical
      scope of an FMI.
      An FMI could support ATM transaction settlement between
      digital pounds and commercial bank money (see row C4), 
      with minimal change to ATM network schemes. \\

    \end{longtable}

  \endgroup
      
\end{center}
\end{landscape}

%%%%%%%%%%%%%%%%%%%
%% (End): Long Table
%%%%%%%%%%%%%%%%%%%

\noindent
We draw the following initial insights (from Figure \ref{fig:uk-cbdc-cap-heatmap}
and Table \ref{table:tab-des-opts-analysis}) 
regarding achieving functional consistency via the various design
options.

\begin{itemize}

  \item Each design
  option's suitability for providing key capabilities by itself
  is summarised below:
  
    \begin{itemize}
      \item \changess{\emph{Option 1. Provided by central bank:}} Although the CBDC system operated by the Bank of
        England would provide foundational
        capabilities
        (such as payments between digital pound wallets)
        as per the platform model,
        it could be unsuitable or partially suitable for
        providing the
        key capabilities identified in this paper.
        This is primarily because providing some of
        the key capabilities could result in users' PII
        becoming accessible to the Bank of England, which
        could potentially conflict with the Bank of England's
        privacy requirements
        and technology design considerations for the digital
        pound (as described in the TWP).
      
      \item \changess{\emph{Option 2. Provided by PIPs:}}
        PIPs could be suitable for providing some of the key
        capabilities (e.g. completing necessary KYC checks).
        Some PIPs may need to partner with existing financial
        institutions to provide certain capabilities
        (e.g. card issuance and transaction processing),
        but they may incur higher costs as a result. 

      \item \changess{\emph{Option 3. Provided by TSPs:}}
        TSPs could be suitable for facilitating confidential
        transfer of payment information 
        between PIPs, 
        particularly if the information is sent \changess{via} direct peer-to-peer channels between PIPs \changess{with}
        the TSPs merely \changess{simplifying}
        onboarding and integration between PIPs.

      \item \changess{\emph{Option 4. Provided by FMI}}
        An FMI may be the most suitable option for
        providing certain key capabilities 
        (e.g. interoperability and programmability).
        This is primarily because a regulated FMI
        can handle PII, become a member of payment schemes,
        and provide common ecosystem services across both
        the digital pound and commercial bank money.

      \end{itemize}

  \item For certain key capabilities, there is no single
    design option that is suitable and, instead, a combination of
    design options may be suitable:

    \begin{itemize}
      \item PIPs and TSPs, working together, could be suitable
        for providing certain key capabilities such as alias
        management
        (with PIPs managing aliases and TSPs
        simplifying \changess{access for ecosystem participants that need to} lookup and
        validate aliases).

      \item PIPs and an FMI, working together, could be suitable
        for providing certain key capabilities such as integrating
        with card schemes and ATM network schemes 
        (with PIPs handling scheme membership and transaction
        processing and the FMI providing interoperability between
        digital pounds and commercial bank money).

    \end{itemize}

  \item Common ecosystem services provided by TSPs and an FMI
    may be needed to enable most of the key capabilities.
    The TSPs and the FMI could also provide these key capabilities
    in a 
    functionally-consistent manner across all forms of 
    retail digital money. 
  
\end{itemize}

%\vspace{-15mm}

%%%%%%%%%%%%%%%%%%%%%%%%%%%%%%%%%%%%%%%%%%%%%%%%%%%%%%%%%%%%%%%%%%%
%% Section 7: Summary and Conclusions
%%%%%%%%%%%%%%%%%%%%%%%%%%%%%%%%%%%%%%%%%%%%%%%%%%%%%%%%%%%%%%%%%%%
\section{Summary and Conclusions}
\label{sec:conclusions}

\changess{In this paper we focused on a potential UK retail CBDC
(the digital pound),
the Bank of England's platform model and}
the risk of fragmentation in payments markets
and retail deposits if digital pounds and commercial bank
money do not have common operational characteristics.
We explored the important concept of functional consistency
as a means to mitigate the risk of fragmentation\changess{,} and we
evaluated design options to support functional consistency
across digital pounds and commercial bank money.
\changess{Our exploration included}

\begin{itemize}

  \item defining functional consistency for money and
    identifying the common
    operational characteristics required to achieve
    functional consistency across all forms of
    regulated retail digital money,

  \item presenting design options (based on the provision
    of these common operational characteristics by the central
    bank, PIPs, TSPs or an FMI) to support functional
    consistency across digital pounds and commercial bank money,

  \item identifying architecturally significant use cases 
    (based on two illustrative user journeys),

  \item selecting the key capabilities for the architecturally-significant
    use cases and identifying the common operational characteristics
    that the key capabilities support,
  
  \item presenting a preliminary evaluation of
    the suitability of the design options
    to provide the key capabilities, and
  
  \item drawing initial insights from the preliminary
    evaluation.

\end{itemize}

\noindent
Our key contribution to the design space for the digital pound
is the insight
that no single design option
could provide all the key capabilities needed for
functional consistency across the digital pound and commercial bank
money.
Instead, a complete solution would need to combine the
suitable design option(s) for delivering each key capability.
This will, therefore, be a complex design exercise requiring
further analysis
and experimentation.

%%%%%%%%%%%%%%%%%%%%%%%%%%%%%%%%%%%%%%%%%%%%%%%%%%%%%%%%%%%%%%%%%%%
%% Section: Further work
%%%%%%%%%%%%%%%%%%%%%%%%%%%%%%%%%%%%%%%%%%%%%%%%%%%%%%%%%%%%%%%%%%%
\section{Further Work}
\label{sec:summary-and-further-work}

Ongoing industry collaboration is needed to 
address the risk of fragmentation in retail deposits
and payments, including evaluating the importance of functional
consistency across digital pounds and commercial bank money
as a means to mitigate the fragmentation risk.

\changess{The analysis framework and common operational
characteristics in this paper could potentially be
used to extend functional consistency to emerging forms of
private digital money such as tokenised deposits and
regulated stablecoins, and to
explore functional consistency for other currencies.
This could include
identifying appropriate architecturally significant use cases,
design options and key capabilities, and evaluating
the suitability of each design option to provide each key
capability.}

\changess{The next steps to validate our initial insights could
also include elaborating an industry architecture with
common ecosystem services provided by an FMI and TSPs,
building prototypes, and
analysing legal, operational and commercial
considerations.
For example, the UK Regulated Liability Network's (RLN)
experimentation phase includes
a common ``platform for innovation'' to
explore functional consistency across digital pounds and
commercial bank money \cite{ukrln-disc-report}.}

We also welcome and encourage the development of novel
functionalities \changess{that} are replicable across all forms of
regulated retail digital money.
We hope this design paper will inspire industry
collaboration on functional consistency during
the design
phase of the digital pound, and we look forward
to feedback.

\vspace{5mm}

\noindent \textbf{Acknowledgements:} 
\noindent
We would like to thank Dincer Ay (Barclays) and Michael Forrest
(Barclays) for helpful input on merchant gateways, card schemes, 
digital identity and Open Banking.

% Return to 'normal' rules on hyphenation for bibliography
\pretolerance=-1
\tolerance=-1
\emergencystretch=0pt

\bibliography{uk-cbdc-design-paper}
\bibliographystyle{plain}

\end{document}